\documentclass[aps,prb,twocolumn,superscriptaddress]{revtex4}
\usepackage{epsfig}
\begin{document}
\title{Signature of Electronic Correlations in the Optical Conductivity \\
of the Doped Semiconductor Si:P}

\author{Marco Hering}
\affiliation{1.~Physikalisches Institut, Universit\"at Stuttgart,
D-70550 Stuttgart, Germany} \affiliation{Physikalisches Institut,
Universit\"at Karlsruhe, D-76128 Karlsruhe, Germany}

\author{Marc Scheffler}
\altaffiliation{Present address: Kavli Institute of Nanoscience, Delft
University of Technology, 2600 GA Delft, The Netherlands}
\affiliation{1.~Physikalisches Institut, Universit\"at Stuttgart,
D-70550 Stuttgart, Germany}

\author{Martin Dressel}
\email[Electronic mail:~]{dressel@pi1.physik.uni-stuttgart.de}
\homepage[URL:~] {http://www.pi1.physik.uni-stuttgart.de}
\affiliation{1.~Physikalisches Institut, Universit\"at Stuttgart,
D-70550 Stuttgart, Germany}

\author{Hilbert v.\ L\"ohneysen}
\affiliation{Physikalisches Institut, Universit\"at Karlsruhe, D-76128
Karlsruhe, Germany}

\date{\today}

\begin{abstract}
Electronic transport in highly doped but still insulating silicon at low temperatures is dominated by hopping between
localized states; it serves as a model system of a disordered
solid for which the electronic interaction can be investigated. We have
studied the frequency-dependent conductivity of phosphorus-doped silicon in the THz
frequency range (30~GHz to 3~THz) at low temperatures $T\geq 1.8$~K.
The crossover in the optical conductivity from a linear to a quadratic
frequency dependence as predicted by Efros and Shklovskii is observed
qualitatively; however, the simple model does not lead to a
quantitative agreement. Covering a large range of donor concentration,
our temperature- and frequency-dependent investigations reveal that
electronic correlation effects between the localized states play an
important and complex role at low temperatures. In particular we find a
super-linear frequency dependence of the conductivity that highlights
the influence of the density of states, i.e.\ the Coulomb gap, on the
optical conductivity.
When approaching the metal-to-insulator transition by increasing doping
concentration, the dielectric constant and the localization length
exhibit critical behavior.
\end{abstract}


\pacs{71.30.+h, 71.45.Gm, 72.15.Rn, 72.20.Ee, 72.40.+w, 72.80.Ng}

\maketitle

\section{Introduction}

The transition from a semiconductor to a metal upon doping is so basic
and important that it seems  hard to believe that this crucial issue is
far from being understood, in spite of the enormous effort invested
over the years.\cite{Lohneysen98,Sarachik95} For low concentration, the
dopants provide localized states which are dispersed in the
semiconductor crystal. When the host contribution to charge transport
can be neglected (at low temperature), the system therefore resembles a
perfectly disordered solid. At low excitation energies the
charge-carrier transport takes place via hopping processes between
these spatially randomly distributed states and is described by
variable range hopping (VRH). Although the model originally suggested
by Mott \cite{Mott1968,MottDavis79} neglects correlation effects
between the localized electronic states, in certain regimes it
describes the phenomena of temperature- and frequency-dependent
transport quite well. Pollak\cite{Pollak70} and
Srinivasan,\cite{Srinivasan71} and later in particular Efros and
Shklovskii (ES) \cite{Efros75} went one step further and included the
long-range electron-electron interaction. During the last thirty years,
a large number of analytical, numerical, and experimental results have
been collected\cite{Efros84,Shklovskii84} which clarified the
importance of Coulomb correlations to some extent. Nevertheless, a
complete and consistent picture cannot be drawn at this point.

In a more general view, the disordered localized states can be called
electron glass: in this terminology Mott's model describes the Fermi glass
\cite{Anderson70} of non-interacting particles, whereas the model taking into account Coulomb
interactions between the states is commonly referred to as the Coulomb
glass.\cite{Davies82}
With decreasing temperature or decreasing electron density, a transition
is expected from a temperature variation of the resistivity
corresponding to that suggested for a Fermi glass to a behavior as
predicted by ES for the Coulomb glass.\cite{Shklovskii84}
The dc conductivity is influenced by a Coulomb
gap $\Delta$ which opens in the density of states
around the Fermi edge $E_F$ due to the long-range Coulomb interaction.
The transition from one behavior to
the other shows that at higher excitation energies the effects of
Coulomb interaction are negligible and Mott's VRH behavior is
recovered.

For the ac conductivity there exist distinct predictions as well. While
Mott described the Fermi glass behavior of non-interacting
particles,\cite{MottDavis79} ES supplemented this derivation by
introducing the interaction between those two states directly involved
in the hopping process that is triggered by resonant absorption of a
photon.\cite{Shklovskii81} Thus with increasing frequency there is also a
transition from the Coulomb glass regime to the Fermi glass.

Within the model of ES,\cite{Shklovskii84} the Coulomb gap can be
roughly evaluated from the temperature dependence of the dc
conductivity in the two regimes. A more direct observation of the
Coulomb correlation gap in the density of states is possible by
tunnelling spectroscopy.\cite{Massey95,Massey00} Quite recently, a
controversy arose whether it is really the Coulomb gap which determines
the crossover energy scale from the Fermi- to the Coulomb-glass-like
behavior in the frequency-dependent transport. According to ES, this
transition is driven by the Coulomb energy of the sites forming the
resonant pair. However, Lee and Stutz\-mann\cite{Lee2001} claimed
instead that the transition occurs when the photon energy equals twice
the Coulomb gap width. A subsequent experimental
study\cite{Helgren2002} and computer simulations\cite{Basylko04} seem
to support the former view. It was recently suggested \cite{Muller04}
that many-particle correlations might be crucial for a quantitative
description of the hopping conductivity. Here we try to resolve this
issue by providing more accurate data in a wider range of carrier
concentration and a broader range of frequency.

A prime example to study the physics of disordered solids is
phosphorus-doped silicon which can be tuned in the insulating regime as
well as through the metal-insulator transition (MIT) at a critical
concentration $n_{c}=3.5 \cdot 10^{18}$~cm$^{-3}$ by varying the donor
concentration $n$ in the single-crystalline
host.\cite{Thomas85,Stupp1993,Lohneysen98} When going to low enough
temperatures and frequencies, the crossover from a Fermi glass to a
Coulomb glass can nicely be investigated in Si:P.\cite{Hornung00} Other
options to tune the system are the application of magnetic
field\cite{Lohneysen91,Lakner94} and
pressure\cite{Paalanen83,Waffenschmidt99} as has also been done for
Si:B.\cite{Bogdanovich99} For small enough doping concentration
$n<2.78\cdot 10^{18}~{\rm cm}^{-3}$, a different effect was inferred by
the strong $\rho(T)$ dependence; the VRH subsides and a simple
activated behavior is observed.\cite{Lohneysen98,Hornung00} Due to
on-site Coulomb repulsion, a Hubbard gap $E_2$ splits the impurity
band. While a clear distinction between uncompensated Si:P and
compensated Si:(P,B) is seen in dc transport,\cite{Liu96} it is not
clear how this effects the frequency-dependent transport.

The work presented in the following mainly concentrates on the
temperature and frequency variation of the conductivity; the
experimental data are compared with the theory by Shklovskii and Efros
in order to elucidate the influence of electronic correlations on the
transport in disordered electron solids. In addition, the measurements
on samples with various doping concentrations give insights into the
scaling behavior at the MIT.

\section{Theoretical considerations}\label{sec:theory}

Before we describe our experimental results, we briefly present the
relevant theories and phenomenological models. More details and
discussion can be found in well established monographs and
reviews.\cite{MottDavis79,Shklovskii84,Castner1991}

\subsection{Fermi glass}

Following Mott's theory of the Fermi glass,\cite{MottDavis79} the
electrical dc conductivity $\sigma_{dc}$ as a function of temperature
$T$ for a non-interacting three-dimensional electron glass (variable range hopping) is
given by the well-known $T^{-1/4}$ law
\begin{equation}\label{EqMottDC}
\sigma_{dc}(T)\propto\exp\{-(T_{\rm Mott}/T)^{1/4}\} \label{eq:dc1}
\end{equation}
with a characteristic temperature
\begin{equation}
T_{\rm Mott}=21.2\frac{1}{k_B\, N_0(E_F)\,\xi^3} \quad . \label{eq:dc2}
\end{equation}
Here $\xi$ is the localization length and $N_0$ is the density of
electronic states (DOS) at the Fermi level $E_F$  which is assumed to
be constant close to $E_F$ in the absence of electronic correlations.
The Mott temperature $T_{\rm Mott}$ tends to zero as the MIT is
approached.\cite{Hornung00}

The common theoretical approach to the ac transport in statistically
disordered solids is the so-called pair approximation. Each
photon-induced hopping process between two localized states has a
certain probability. The sum over all the hops contributing to the
absorption process leads to the photon-assisted conductivity of the
system. In the derivation of the ac conductivity, a simple one-electron
model of the disordered system is assumed, neglecting any correlation
effects between the states.\cite{MottDavis79} Apart from logarithmic
corrections, the frequency-dependent ($\omega$=2$\pi f$) conductivity
of a Fermi glass at $T=0$ shows a quadratic behavior:
\begin{equation}\label{EqMott}
\sigma_1(\omega) = \alpha e^2 N_0^2\, \hbar \omega^2 \xi\, r_{\omega}^4 \quad.
\end{equation}
Here we denote by $\sigma_1$ the real part of the complex conductivity
$\hat{\sigma}=\sigma_1+i\sigma_2$; the imaginary part is related to the
real part of the dielectric constant
$\epsilon_1=1-4\pi\sigma_2/\omega$.\cite{DresselGruner02}
The length $r_\omega = \xi \cdot \ln\{2 I_0 / \hbar \omega\}$ is the most
probable distance for the hopping processes. Here $I_0$ is the
prefactor describing the overlap integral; its value is of the order of
the binding energy of the electronic state,\cite{Castner1991} and hence
it is commonly taken to be the Rydberg energy of the dopant (for Si:P we
use $I_0 \approx 45$~meV). The factor $\alpha$ is a constant of order
of unity.

\subsection{Coulomb glass}

In the case of appreciable (long-range) electronic interaction, the DOS approaching
the Fermi energy is reduced due to the Coulomb interaction between the
localized states, leading to $N(E)\approx (3/\pi)(E-E_F)^2
\epsilon_1^3/e^6$. The width of this so-called Coulomb gap is given by
\begin{equation}
\Delta \propto {e^3N_0^{1/2}}/{\epsilon_1^{3/2}} \quad.
\label{eq:Delta}
\end{equation}
This concept was confirmed by tunnelling
exper\-i\-ments\cite{Massey95,Massey00} which reveal a parabolic
dependence of the conductance measuring the DOS. The typical size of
the Coulomb gap in doped semiconductors is 1~meV. Taking the Coulomb
gap into account, Shklovskii and Efros\cite{Shklovskii84} calculated
the temperature dependence of the dc conductivity, the so-called
$T^{-1/2}$ law
\begin{equation}\label{EqESDC}
\sigma_{dc}(T)\propto\exp\{-(T_{\rm ES}/T)^{1/2}\} \label{eq:dc3}
\end{equation}
with the characteristic temperature
\begin{equation}
T_{\rm ES}=2.8\frac{e^2}{k_B\,\epsilon_1\,\xi} \quad , \label{eq:dc4}
\end{equation}
which is a measure of the characteristic Hartree interaction strength.
At large enough $T$ the accessible energy range for hopping processes,
$k_B(T^3T_{\rm Mott})^{1/4}$, exceeds the Coulomb gap $\Delta$, and
Mott's VRH behavior Eq.\ (\ref{eq:dc1}) is recovered. This crossover
from a Fermi glass to a Coulomb glass can be seen as a weak kink in the
temperature-dependent conductivity.\cite{Hornung00}

The ac conductivity is again calculated in the pair approximation. When
taking into account the average Coulomb attraction
$U(r_{\omega})=e^2/\epsilon_1 r_{\omega}$ between two sites forming a
resonant pair of distance $r_{\omega}$, hopping processes can even
occur between states that are separated energetically by more than the
photon excitation energy $\hbar \omega$. This long-range Coulomb
interaction $U(r_{\omega})$ strongly depends on the real part of the
dielectric constant $\epsilon_1$. As will be seen later, $\epsilon_1$
diverges as the MIT is approached; it basically screens the Coulomb
interaction. The resulting optical conductivity
\begin{equation}\label{EqES1}
  \sigma_1(\omega) = \alpha e^2 N_0^2\, \omega \,\xi\, r_\omega^4\, [\hbar \omega + U(r_{\omega})]
\end{equation}
is distinct from Eq.\ (\ref{EqMott}) for small excitation energies
$\hbar \omega \ll U(r_{\omega})$, when the frequency-dependent
conductivity yields approximately a linear behavior:
\begin{equation}\label{EqES2}
  \sigma_1(\omega) = \frac{\alpha e^4}{\epsilon_1} N_0^2\, \hbar \omega \, \xi^4\, \left[\ln\{2 I_0/\hbar \omega\}\right]^3 \quad .
\end{equation}
In the opposite case of weak correlations, $\hbar \omega \gg
U(r_{\omega})$, the frequency dependence of Eq.\ (\ref{EqES1}) is
quadratic if the logarithmic corrections are neglected. Naturally, it
corresponds to Mott's formula Eq. (\ref{EqMott}). Hence the transition from
linear to quadratic behavior can be considered as a transition from probing the
interacting Coulomb glass at low frequency to the Fermi glass regime at high frequency where correlation
effects can be neglected.

Although for both the temperature-dependent dc data and the
low-temperature ac behavior a crossover is predicted, the canonical
theory presented so far does not represent a direct correspondence. In
the former case the reduced DOS due to Coulomb interaction -- the
Coulomb gap -- modifies Mott's $T^{-1/4}$ law to the ES $T^{-1/2}$ law.
In the latter case, however, Eq.\ (\ref{EqES1}) and Eq.\ (\ref{EqES2})
have been calculated under the assumption of a constant DOS at the
Fermi level although meant to describe the interacting system. Since
the DOS of a Coulomb glass exhibits the Coulomb gap of width $\Delta$,
this assumption is valid only for $U(r_{\omega}) \gg \Delta$, i.e.\
when the Coulomb interaction influences the occupation numbers only to
a small extent. Then only states outside the gap contribute to
$\sigma_1(\omega)$.

In the opposite case, mainly states within the gap participate in the
hopping conduction. For $\Delta > U(r_{\omega}) > \hbar \omega$,
Shklovskii and Efros\cite{Shklovskii81} derive a stronger frequency
variation than given by Eq.\ (\ref{EqES2}). Hence, the incorporation
of the Coulomb gap results in a super-linear frequency dependence
\begin{equation}\label{EqES3}
  \sigma_1(\omega) = \frac{\alpha e^4}{\epsilon_1} N_0^2\,  \xi^4 \frac{\hbar \omega}{\ln\{2 I_0/\hbar \omega\}}
\end{equation}
in contrast to the sub-linear dependence of Eq.\ (\ref{EqES2}) for
constant DOS. Between both limiting cases a smooth crossover is
expected. Computer simulations\cite{Basylko04} yield larger exponents
and a more abrupt transition compared to the predictions of
ES.\cite{Shklovskii81,Shklovskii84}

In principle, frequency-dependent measurements provide the possibility
to investigate electron glasses like Si:P in the relevant energy
scales. But due to experimental difficulties to access the very low
energies required (meV and less, corresponding to the GHz and THz
frequency range), the expected crossover from Coulomb glass to Fermi
glass has been observed only recently.\cite{Lee2001,Helgren2002} At
this point it is not clear how the opening of a Hubbard gap in the
density of states influences the frequency dependent conductivity.

\subsection{Metal-insulator transition}

So far we have only discussed the insulating state of disordered
systems. In our case the doping of the host silicon crystal adds an
extra dimension as we approach the MIT with increasing doping. This
quantum phase transition from an insulator with localized states and
zero dc conductivity at $T=0$ to a metal with extended electronic
states associated with finite dc conductivity can be driven by varying
external parameters like uniaxial stress, magnetic fields or, in our
case, the doping concentration
$n$.\cite{Thomas85,Belitz94,Sondhi97,Lohneysen98}  The MIT
in doped semiconductors like Si:P is due to two facts: the localization
arising from disorder (Anderson transition) and from long-range
electron-electron interaction due to the loss of screening as states
become localized (Mott-Hubbard transition).\cite{Lee85,Belitz94,Lohneysen98} The
spatial extent of the electronic states scales with doping;
the localization length increases as the MIT is approached
\begin{equation}
\xi\propto|n_c-n|^{-\nu}
\quad , \label{eq:nu}
\end{equation}
where $n_c$ is the critical
concentration.\cite{Shklovskii84,Anderson58,Wegner76,Stauffer79} On the metallic side also the conductivity
scales:\cite{Webman75,Efros76} $\sigma_{1} \propto\linebreak |n_c-n|^{\mu}$.
Theoretically, $\mu$ is inferred from the correlation-length
critical exponent $\mu=(d-2)\nu$ by scaling arguments;\cite{Wegner76}
$d$ indicates the dimensionality of the system. Also the dielectric
constant diverges as\cite{Bergman77,Efros76}
\begin{equation}
\epsilon_1 \propto |n_c-n|^{-\zeta^{\prime}}
\quad . \label{eq:zetaprime}
\end{equation}
The power laws are supposed to be universal; that is the critical
exponents do not depend on the details of the geometric structure or the interaction.
Previous investigations of uncompensated semiconductors inferred an
exponent $\mu=0.5$, in contrast to $\mu=1$ found for compensated
semiconductors and amorphous metals.\cite{Thomas85} However, it turned
out that the exponent depends on the parameter by which the MIT is
tuned and on how close to the MIT the experiments are
performed.\cite{Lohneysen98} Numerical simulations and experiments on
metal-insulator composites give values for the critical exponent that
scatter considerably, for instance $\zeta^{\prime}$ between 0.5 and 1.
\cite{Grannan81,Song86,Clarkson88,Brouers91} In three dimensions, the
relation $\zeta^{\prime}=2\mu$ is expected.\cite{Imry82}
McMillan\cite{McMillan1981} suggested that both critical exponents are
related by $\zeta^{\prime} = \nu (\eta -1)$ where $1 < \eta < 3$.
Applied to our case, it is not obvious that the crucial assumption of
this model really holds that only states within the Coulomb gap are of
relevance. Recent computer calculations support this
scepticism.\cite{Basylko04}

The above considerations neglect the on-site Coulomb interaction: For
uncompensated semiconductors like Si:P, a twofold spin-degenerate ground
state 1s (A$_1$) (and higher valley-orbit split states not considered
here) is induced with each P donor atom. Hence uncompensated Si:P is
always at half-filling of the impurity band; only the lower
Hubbard band is occupied. With increasing $n$ the two Hubbard bands
start overlapping at $n_0=0.8 n_c$ \cite{Liu96,Lohneysen98} but states
at the edges are still Anderson localized. It is only at $n_c$ that
they become extended.

\section{Experiments and Results}

\begin{table*}
\caption{\label{tableofsamples} List of Si:P crystals used in this
study. The phosphorus concentration is indicated by $n$; also the ratio
with respect to the critical concentration
$n_c=3.5\cdot10^{18}$~cm$^{-3}$ is given. $\rho(300~{\rm K})$ is the dc
resistivity measured at room temperature; the low-temperature to
high-temperature resistivity ratio is indicated by $\rho(4.2~{\rm
K})/\rho(300~{\rm K})$. The crossover frequency $\omega_{c1}/2\pi$ is
obtained by the intersection of the linear and quadratic fits of the two
regimes. If the frequency-dependent conductivity is fitted by power
laws $\sigma_1(\omega)\propto\omega^s$, the exponents $s_{\rm CG}$ and
$s_{\rm FG}$ are obtained for Coulomb glass and Fermi glass,
respectively. These fits give the approximate crossover frequency
$\omega_{c2}/2\pi$ between both regimes. For samples marked
$\triangleright$ ($\triangleleft$) the conductivity $\sigma_1$ could not
be determined from Fabry-Perot resonances because the absorption is too
high (low). For sample No.\ 5 only the dielectric constant was measured at low temperatures but not the conductivity (denoted by -). $T_0$ is the temperature of the minimum in the dielectric
constant. $\epsilon_1$ corresponds to the dielectric constant measured
at $T=1.8$~K in the THz range of frequency; $\chi$ is the dielectric
susceptibility.}
\begin{tabular}{|c|cc|cc|c|ccc|cc|c|}
  \hline
  No. &  $n$ & $n/n_c$ & $\rho(300~{\rm K})$ & $\frac{\rho(4.2~{\rm K})}{\rho(300~{\rm K})}$&  $\omega_{c1}/2\pi$ & $s_{\rm CG}$ & $s_{\rm FG}$ & $\omega_{c2}/2\pi$ & $T_0$& $\epsilon_1$ & $4\pi \chi$ \\
& ($10^{18}$~cm$^{-3}$) & & ($\Omega$cm) &  & (GHz) & & & (GHz) & (K) & & \\
  \hline
  1 & 0.89 & 0.25 & 0.0245 & $4.4 \cdot 10^{10}$ & $\triangleleft$ & $\triangleleft$& $\triangleleft$ & $\triangleleft$& 11 & 12.90 & 1.20 \\
  2 & 1.60 & 0.46 & 0.0180 & $1.48 \cdot 10^8$ & 540    & 1.16 & 2.15 &  760   & 9  & 15.51 & 3.81 \\
  3 & 1.97 & 0.56 & 0.0162 & $1.53\cdot 10^7$  & 460    & 1.21 & 2.16 &  550   & 8  & 18.13 & 6.43 \\
  4 & 2.29 & 0.65 & 0.0149 & $3.49\cdot 10^6$  & 480    & 1.17 & 2.25 &  630   & 6  & 20.47 & 8.77 \\
  5 & 2.40 & 0.69 & 0.0145 &          -        &    -    &   -   &   -   &    -    & 6  & 23.56 & 11.86 \\
  6 & 2.57 & 0.73 & 0.0139 & $3.46\cdot 10^4$  & 440    & 1.33 & 2.28 &  570   & 5  & 26.48 & 14.78 \\
  7 & 2.91 & 0.83 & 0.0130 & 1060 & $\triangleright$ & $\triangleright$ & $\triangleright$& $\triangleright$ & 5  & 41.57 & 29.87 \\
  8 & 3.04 & 0.87 & 0.0127 & 80 & $\triangleright$ & $\triangleright$& $\triangleright$ &$\triangleright$ &    & 55.45 & 43.75 \\
    \hline
\end{tabular}
\end{table*}

The samples for this study were cut from two different Czochralski-grown,
nominally uncompensated silicon single crystals with a
phosphorus gradient along the growth axis.\cite{remark4} From the
ingots a number of disks were prepared using a diamond wire saw. To
remove distorted surface layers,\cite{Outuka80} the crystals were
chemically and mechanically treated by well established procedures.
According to Thurber {\it et al.} \cite{Thurber80} the doping
concentration was determined from the room-temperature resistivity
employing a commercial four probe measurement system (FPP 5000 by Veeco
Instruments). For high doping levels the resistivity ratio
$\rho(4.2~{\rm K})/\rho(300~{\rm K})$ (determined from standard dc
measurements) is consistent with Ref.\ \onlinecite{Hornung93}.
The dopant concentration of the crystals used in this study ranges
between $0.89\cdot10^{18}$ cm$^{-3}$ and $3.04\cdot10^{18}$ cm$^{-3}$;
the sample properties are summarized in Table~\ref{tableofsamples}.

For optical experiments specimens of $10\times 10$~mm$^2$ were prepared
with different thickness ranging from 2~mm to 0.05~mm and less. The
crystals were chemically and mechanically polished to optical quality.
In combination with highly parallel opposite faces, this ensures
pronounced Fabry-Perot resonances.
Using a quasi-optical Mach-Zehnder interferometer in the THz range of
frequency equipped with backward-wave oscillators as coherent and
tunable radiation sources,\cite{Kozlov1998} the optical transmission
and change in phase could be measured between 30~GHz and 1.2~THz,
corresponding to an energy range between 0.1~meV and 5~meV.
The Fabry-Perot resonances due to multi-reflection at the surfaces of
the silicon sample were utilized to enhance the sensitivity and
accuracy.\cite{remark1}
From the spectra the real parts of the conductivity
$\sigma_{1}(\omega)$ and the dielectric constant $\epsilon_{1}(\omega)$
can be calculated using the Fresnel formula\cite{DresselGruner02} with
an uncertainty of 10\% and 5\%, respectively. Complementary optical
experiments were performed using  a Fourier transform spectrometer
(modified Bruker IFS 113v) up to frequencies of 3~THz (corresponding to
12~meV).
Depending on the dopant concentration, two to four different thicknesses
were measured of each sample in order to optimize the sensitivity. For
the sample with phosphorus concentration $n=0.89\cdot10^{18}$cm$^{-3}$,
the absorption is too weak to determine
the conductivity in the entire frequency range.
On the other hand, highly conducting specimens with
$n\geq2.9\cdot10^{18}$ cm$^{-3}$ do not transmit sufficiently
to achieve a signal-to-noise ratio large enough to analyze the data
with respect to $\sigma_{1}$ over the whole accessible frequency range.

\begin{figure}
\centerline{\includegraphics{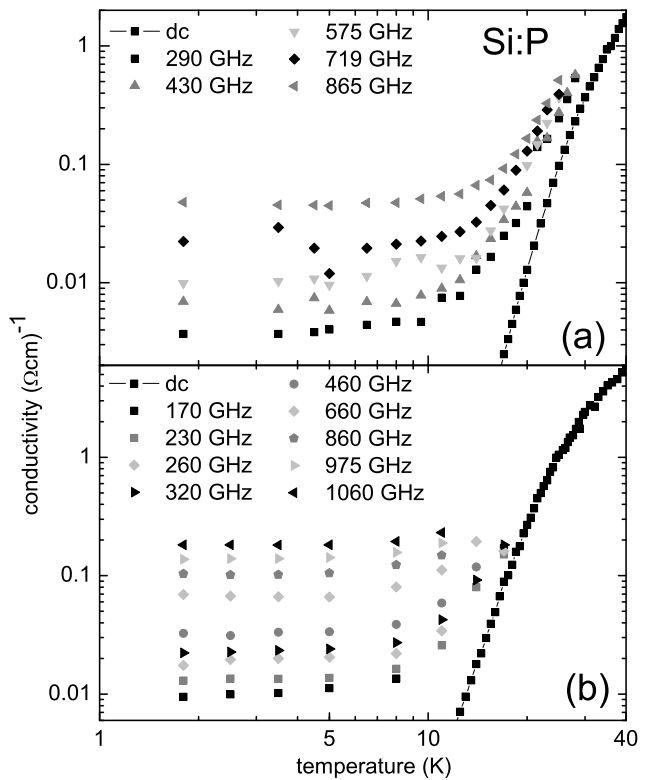}}
\caption{\label{FigTempDepSigma}Temperature dependence of the real part
of the conductivity $\sigma_{1}$ for various frequencies and the
temperature dependent dc results for Si:P crystals with (a) $n=1.6\cdot
10^{18}$~cm$^{-3}$ and (b) $n=1.97\cdot 10^{18}$~cm$^{-3}$.}
\end{figure}
Temperatures down to 1.8~K were reached with a pumped liquid $^4$He
optical cryostat. As an example, in Fig.\ \ref{FigTempDepSigma}(a) the
temperature dependence of the conductivity as obtained from optical
measurements is displayed for a sample with $n=1.6\cdot
10^{18}$~cm$^{-3}$. The ac conductivity curves below approximately 10~K
suggest that phonon-induced charge-carrier transport can be neglected,
and thus the system is in the so-called zero-phonon regime; in other
words $\hbar\omega > k_BT$. This is a basic requirement in order to
compare the measured conductivity spectra $\sigma(\omega)$ with the
theoretical predictions discussed above. Because phonon-induced
processes dominate at higher temperatures, the ac conductivity curves
and the dc data coincide. Obviously, the higher the frequency of a
particular curve, the higher the temperature at which this occurs.
Similar behavior is observed for the other samples. Fig.\
\ref{FigTempDepSigma}(b) shows the data for a crystal with higher
phosphorus concentration $n=1.97\cdot 10^{18}$~cm$^{-3}$.

\section{Analysis and Discussion}

\subsection{AC Conductivity}

\begin{figure}[b]
\centerline{\includegraphics{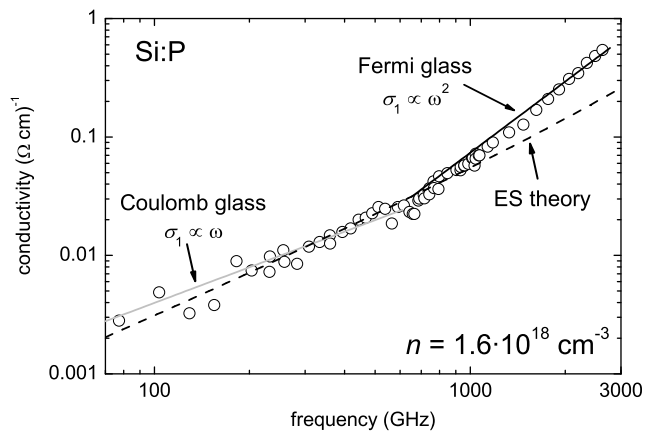}}
\caption{\label{FigFreqDep1}Frequency dependence of a Si:P sample with
$n=1.6\cdot 10^{18}$~cm$^{-3}$ at $T=1.8$~K. The straight lines are
fits with linear and quadratic behavior respectively. The dashed line
is a fit with the prediction of Efros and
Shklovskii\protect\cite{Shklovskii84} given in Eq.\ (\ref{EqES1}). The
error bars are of the size of the symbols.}
\end{figure}

From the frequency-dependent transmission measurements at various
temperatures, the optical conductivity and dielectric constant could be
evaluated for samples No.\ 2 through 6 as described above. This covers
the doping range from $1.6\cdot 10^{18}$~cm$^{-3}$ to $2.57\cdot
10^{18}$~cm$^{-3}$. Here we confine ourselves to the lowest-temperature
data ($T=1.8$~K) in order to ascertain that the experiments are
performed in the zero-phonon regime ($k_BT<\hbar\omega$) for all
frequencies of interest. In Fig.\ \ref{FigFreqDep1} the optical
conductivity for the sample with doping concentration $n=1.6\cdot
10^{18}$~cm$^{-3}$ is presented. The data can be well described by a
crossover from a linear to a quadratic behavior of $\sigma_1(\omega)$
as indicated by the solid lines. As pointed out in
Sec.~\ref{sec:theory}, this corresponds to the limiting cases of a
Fermi glass with no interaction at high frequencies and a Coulomb glass
with electronic correlations at low frequencies, respectively. The
corresponding crossover frequencies $\omega_{\rm c1}$ are listed in
Tab.~\ref{tableofsamples}.

The dashed line in Fig.~\ref{FigFreqDep1} is drawn according to the
quantitative description given in Eq.\ (\ref{EqES1}); obviously the
experimentally observed crossover is much more pronounced than the
smooth transition predicted by the theory of Shklovskii and
Efros.\cite{Shklovskii81,Efros84,Shklovskii84} This strong discrepancy
of theory and experiment has already been observed by previous
investigations.\cite{Lee2001,Helgren2002}

\begin{figure}
\centerline{\includegraphics{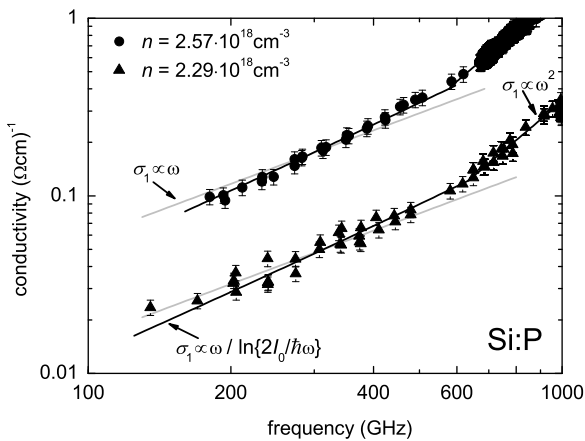}}
\caption{\label{FigFreqDep2}Frequency-dependent conductivity in the
Coulomb glass regime for the silicon samples with phosphorus concentration
$n=2.29\cdot 10^{18}$~cm$^{-3}$ and $n=2.57\cdot 10^{18}$~cm$^{-3}$.
The fits correspond to a super-linear behavior as indicated in the
graph with $I_0=45$~meV. At higher frequencies, the transition to
quadratic behavior is apparent. Linear behavior is plotted as gray
lines for comparison.}
\end{figure}
Upon closer inspection as presented in Fig.\ \ref{FigFreqDep2}, the
low-frequency data of all samples are better described by a
super-linear frequency dependence as suggested by ES in
Eq.\ (\ref{EqES3}) than by a purely linear behavior. (Note that the
logarithmic corrections in Eq.\ (\ref{EqES2}) even lead to a sub-linear
frequency dependence.) The logarithmic corrections $\sigma_{1}(\omega)
\propto \omega/\ln\{2I_{0}/\hbar \omega\}$ for Coulomb glasses have to
be taken into account if the Coulomb interaction energy is smaller than
the width of the Coulomb gap, $U(r_{\omega})< \Delta $. The resulting
fits are shown for two samples in Fig.~\ref{FigFreqDep2}, with the
exactly linear fits for comparison.

\begin{figure}
\centerline{\includegraphics{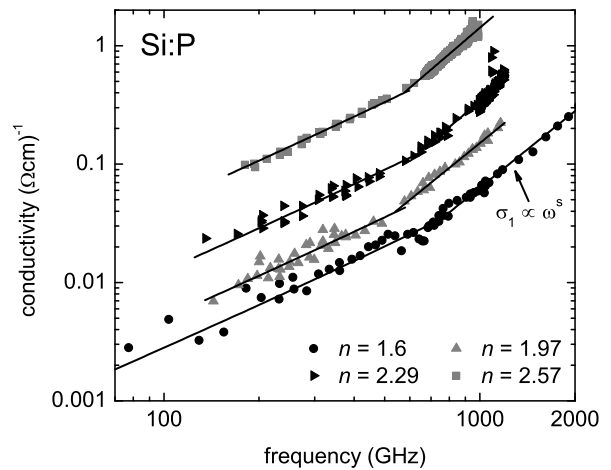}}
\caption{\label{FigFreqDep3}Frequency-dependent conductivity of the
four experimentally accessible Si:P samples. The respective phosphorus
concentration is indicated in the graph in units of $10^{18}$
cm$^{-3}$. The fits correspond to the
super-linear and approximately quadratic behavior.}
\end{figure}

In the opposite limit of high frequencies where the predictions for a
Fermi glass apply, a slightly stronger frequency variation is
recovered than expected from Eq.\ (\ref{EqES1}). A fit by a power-law
$\sigma(\omega) \propto \omega^s$ yields the exponent $s$. As seen in
Tab.~\ref{tableofsamples}, $s(n)$ increases with doping concentration
from $s=2.15$ for $n=1.6\cdot 10^{18}$~cm$^{-3}$ to $s=2.28$ for
$n=2.57\cdot 10^{18}$~cm$^{-3}$. An overview of the spectra of samples
with different phosphorus concentration is given in
Fig.~\ref{FigFreqDep3} together with the corresponding fit of the
super-linear behavior at low frequencies and the approximately
quadratic dependence at higher frequencies.

We may compare our findings to previous investigations deep in the
Coulomb glass state. In the radio-frequency range ($10^2$ to $10^5$~Hz)
at $T=13$~mK (in a sample close to the MIT and under uniaxial stress) Paalanen
{\em et al.}\cite{Paalanen83} found $\sigma_1(\omega)\propto \omega^s$
with $s=0.9$ in Si:P. Castner and coworkers investigated Si:As with
$\omega/2\pi < 2$~GHz for $T>1$~K and found a super-linear frequency
dependence with $1<s< 1.5$ and a very strong doping dependence of $s$
which was interpreted as possible qualitative agreement with Mott's
$\omega^2$ law.\cite{Castner86} For Si:P they observed an increase of
$s(n)$ from about unity to 2.2 with increasing $n$ in a density range
comparable to our samples.\cite{Migliuolo88} Based on previous
experiments,\cite{Hess82} Castner\cite{Castner1991} argues that this
might be due to a decrease in the Coulomb energy $U$ as the MIT is
approached. In the light of our results, the following interpretation
seems plausible: the observed exponents for Si:As correspond to the
super-linear behavior in the Coulomb gap regime, and for Si:P the
concentration range might include the transition from Coulomb- to
Fermi-glass behavior but was not identified because of the limited
number of frequencies in a comparably small frequency range.

As already pointed out above, the transition between the two regimes is
clearly seen and much more abrupt than expected from the general
description in Eq.\ (\ref{EqES1}) suggested by ES. From the intersection
of the fit curves to power-laws, a crossover energy $\hbar \omega_{c2}$
can be defined. The values obtained by this procedure are summarized in
Tab.~\ref{tableofsamples}. It should be noted that the resulting
crossover frequencies $\omega_{c2}$ are larger than the values
$\omega_{c1}$ obtained from the intersection of linear and quadratic
fits (also listed in Tab.~\ref{tableofsamples}).

The crossover frequency is of particular interest because it is a
quantity that can easily be obtained from the experimental data and
furthermore gives a direct measure for the interaction energy $U$:
following Eq.\ (\ref{EqES1}), at the crossover frequency we have $\hbar
\omega_c = U(r_{\omega_c})$. In general one expects the crossover frequency to
decrease with increasing doping, approaching the MIT. This can be explained by the stronger screening of the
Coulomb interaction due to the increase of the dielectric constant, as
discussed in Sec.~\ref{sec:MIT} below.
Helgren {\it et al.}\cite{Helgren2002}
suggested that the crossover frequency scales with the doping
concentration $\omega_c\propto |1- n/n_c|^{\beta}$ with $\beta\approx
1.65$; however, compared to those reports we observe a much weaker
dependence; a power-law fit would yield an exponent $\beta\approx 0.21$. Note, however, that our
values, but in particular the conductivity data and subsequently
$\omega_c$ reported in Ref.~\onlinecite{Helgren2002} have considerably large
error bars; hence we should not over-interpret this discrepancy.

Lee and Stutzmann \cite{Lee2001} suggested that the width of the
Coulomb gap affects the crossover energy. Combining temperature- and
frequency-dependent data on Si:B they observe a relation
$\hbar\omega_{c2}\approx 2\Delta$. Taking our sample with
$n=2.57\cdot10^{18}$~cm$^{-3}$ as an example, the temperature-dependent
resistivity exhibits a kink around 4.6~K which indicates the transition
from the Mott regime to the Efros-Shklovskii regime. From the fits in
both limits [Eqs. (\ref{EqMottDC}) and (\ref{EqESDC})] we obtain
$T_{\rm Mott} =2.25\cdot 10^6$~K and $T_{\rm ES}=790$~K which yields a
Coulomb gap $\Delta=1.15$~meV, following Ref.~\onlinecite{Massey00}.
The corresponding frequency 280~GHz is around half of the crossover
frequency $\omega_c/2\pi=570$~GHz, as suggested in
Ref.~\onlinecite{Lee2001}.

Whether this agreement indicates a close link between the width of the
Coulomb gap and the crossover between the two regimes indicated in ac
transport, however, remains an open question. One reason is that the
validity of Eq.\ (\ref{EqES1}) is unclear for the important and
experimentally relevant case of $U(r_{\omega})\approx 2\Delta$. Lee and
Stutzmann proposed that the sharp crossover is controlled not by the
average interaction strength $U$ as in Eq.\ (\ref{EqES1}), but instead
by the Coulomb gap. In other words, it is not the mean Hartree energy
between sites forming a resonant pair that is relevant, but the
correlation energy $2\Delta$. For this reason the single-particle gap
measured by tunnelling is much larger than the renormalized Coulomb gap
that governs transport. Helgren {\it et al.},\cite{Helgren2002} on the
other hand, suggest that the Coulomb interaction determines the
observed frequency-dependent crossover from ES- to Mott-like hopping
conduction occurs. Based on recent computer
simulations, Basylko {\it et al.}\cite{Basylko04} also argue that the
transition is driven by the Coulomb energy of sites forming resonant
pairs and not by the width of the Coulomb gap. They calculated the
number of sites outside the Coulomb gap relative to  the total number
of sites participating in the ac conductivity at a given frequency
$\omega$, and found that within the frequency range of interest only
sites inside the Coulomb gap contribute to the ac conductance, in
accordance with our observation of a super-linear frequency dependence.

\subsection{Metal-Insulator Transition}
\label{sec:MIT}

\begin{figure}[b]
\centerline{\includegraphics[width=6cm]{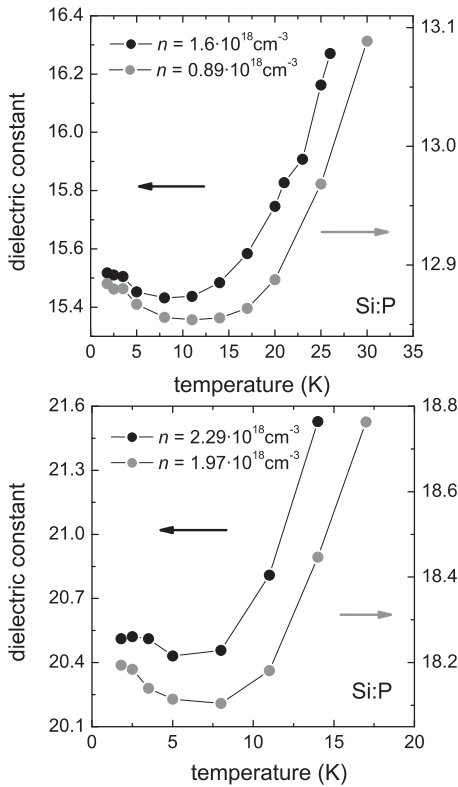}}
\caption{\label{FigTempDepEps1}Temperature dependence of the
dielectric constant $\epsilon_{1}$ for Si:P samples with different
phosphorus concentrations $n$ as indicated. The curves correspond
to the left and right axes as indicated; also note the different
temperature axes for the two frames. The data points are mean
values averaged over the frequency range investigated. }
\end{figure}

Besides the frequency-dependent conductivity, the dielectric constant
$\epsilon_{1}(\omega,T)$ was measured for each Si crystal of different
phosphorus concentration. In the entire frequency range $30~{\rm
GHz}<\omega/2\pi<3$~THz, the dielectric constant $\epsilon_{1}(\omega)$
is basically frequency independent up to about $T=30$~K.

Surprisingly, as presented in Fig.~\ref{FigTempDepEps1}, we find a
non-monotonic temperature dependence $\epsilon_{1}(T)$ with a
pronounced minimum at low temperatures $T_0$ which might be an inherent
property of the zero-phonon regime.\cite{remark2} Depending on the
phosphorus concentration, the minimum decreases from approximately
$T_0=11$~K for $n=0.89\cdot10^{18}$~cm$^{-3}$ to  $T_0=5$~K for samples
with $n=2.91\cdot10^{18}$~cm$^{-3}$. Within our range the concentration
dependence of $T_0$ can be approximated by a linear behavior.
Temperature-dependent measurements on undoped silicon do not exhibit
such a minimum in the dielectric constant.\cite{Tan81} Also for samples
with higher concentration there is no minimum evident in the observed
temperature range; for these samples $T_0$ is very likely below the
lowest temperature of $T=1.8$~K that could be reached with our optical
cryostat. It is tempting to suggest that the opening of the Hubbard gap
$E_2$ leads to an additional contribution to the dielectric constant
$\Delta\epsilon_1 \propto E_2^{-2}$. Similar experiments on compensated
samples could clarify this issue.\cite{remark5}

As can be seen in Fig.~\ref{FigTempDepEps1}, for $T>T_0$ the
dielectric constant increases with temperature because relaxation
processes due to phonons become more likely. Note that the values of the
dielectric constant for each temperature are obtained as the average of
dozens of data points acquired in the frequency regime under
inspection; they are very accurate (better than 0.5~\%) and geometrical
effects can be ruled out.

\begin{figure}
\centerline{\includegraphics[width=7cm]{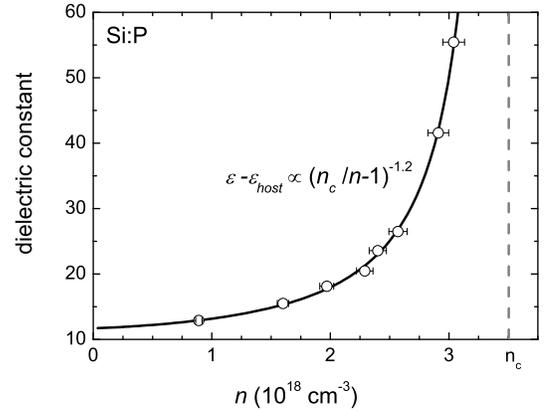}}
\caption{\label{FigConcDepEps2}Dielectric constant $\epsilon_1$ of
Si:P as a function of phosphorus concentration measured at
$T=1.8$~K. The solid line represents a fit by
$\epsilon_{1}-\epsilon_{\rm host}\propto
(n_c/n-1)^{-\zeta^{\prime\prime}}$ with
$\zeta^{\prime\prime}=1.2$.}
\end{figure}

\begin{figure}
\centerline{\includegraphics[width=6.5cm]{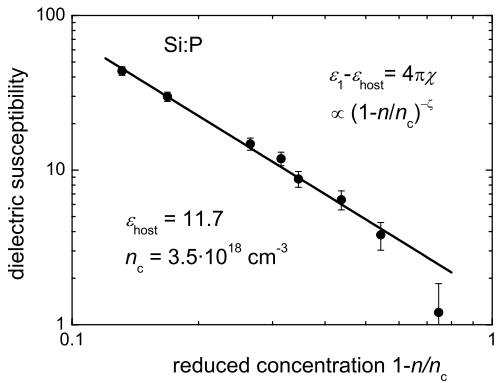}}
\caption{\label{FigConcDepEps}Dielectric susceptibility $4\pi
\chi$ of Si:P crystals versus the reduced doping concentration
$1-n/n_c$ with $n_c=3.5\cdot 10^{18}$~cm$^{-3}$ at $T=1.8$~K.}
\end{figure}

The low-temperature ($T \ll T_0$) dielectric constant $\epsilon_{1}$
diverges when the phosphorus concentration increases towards the MIT
as depicted in Fig.~\ref{FigConcDepEps2}.
In the present case it is more appropriate to subtract the lattice term
$\epsilon_{\rm host}=11.7$ of undoped silicon in order to gain direct insight to the critical behavior of the disordered localized system itself.
Consequently Fig.~\ref{FigConcDepEps} shows the concentration
dependence of the dielectric susceptibility fitted by
\begin{equation}
4\pi\chi = \epsilon_{1}-\epsilon_{\rm host}\propto (1- n/n_c)^{-\zeta}
\quad .\label{eq:zeta}
\end{equation}
The critical exponent $\zeta=1.68$ is in reasonable accord with previous
reports ($\zeta=2.0$, Ref.~\onlinecite{Helgren2002}) taking the error bars of about 20\%\ into account when comparing different samples.
Significant deviations are observed for small phosphorus concentrations, i.e.\ further away from the MIT.

A resonable fit is obtained in the range $0.13 > 1- n/n_c >0.5$. It should be pointed out that there is no clear indication of the opening of the Hubbard gap that was inferred to occur around $1 - n/n_c \approx 0.8$ from transport measurements.\cite{Liu96} We also note that for more heavily doped samples on the metallic side, the critical region of the critical exponent $\mu \approx 1$ of electrical dc conductivity, $\sigma_{dc} \propto \mid n - n_c \mid^{\mu}$, is much smaller, i.e. $ 1 - n/n_c \approx 0.2$.\cite{Stupp1993}

There is some confusion in the literature because of an apparent ``scaling'' with the variable $n_c/n -1$, i.e.\ $\chi \propto |n_c/n - 1|^{- \zeta^{''}}$, extending
to much larger reduced concentrations.\cite{Hess82,Capizzi80} However, there is no theoretical justification for such a ``scaling''. For demonstration purpose only we note that the line shown in Fig.~\ref{FigConcDepEps2} is actually a fit with this function, yielding $\zeta^{''} = 1.2$ in strong contrast to $\zeta = 1.68$, obtained from the fit to Eq.~(\protect\ref{eq:zeta}).

\begin{figure}
\centerline{\includegraphics[width=7cm]{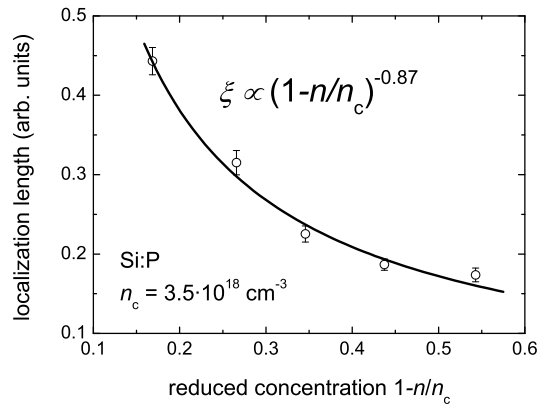}}
\caption{\label{FigConcDepLoc}Scaling behavior of the localization
length in Si:P. The critical concentration is $n_c=3.5\cdot
10^{18}$~cm$^{-3}$.}
\end{figure}

From temperature-dependent dc measurements Hornung
{\it et al.}\cite{Hornung00} evaluated the concentration dependence of
the Mott temperature $T_{\rm Mott}$ and via Eq.\ (\ref{eq:dc2}) the
scaling exponent $\nu=1.1$ for  the correlation length. A scaling
behavior is expected only very close to the MIT, nevertheless we can
extract the relative concentration dependence of the localization
length from the frequency-independent prefactor in Eq.\ (\ref{EqES3}),
if we take the low-temperature value of the conductivity and dielectric
constant. These results, plotted in Fig.~\ref{FigConcDepLoc}, can be
well described by a critical exponent of $\nu=0.87$ according to Eq.\
(\ref{eq:nu}). Our findings are in excellent agreement with similar
experiments for which $\nu=0.83$ was reported.\cite{Helgren2002} They
fulfill the Harris criterion\cite{Harris74} $\nu > 2/d$, with $d$ the
spatial dimension, which implies that the disorder does not affect the
critical behavior.

Castner and coworkers\cite{Migliuolo88,Castner1991} suggested a
relation of the dielectric susceptibility to DOS and to the
localization length
\begin{equation}
\chi= e^2\; N(E_F)\; \xi^2 \quad .
\end{equation}
Thus the ratio of both critical exponents $\zeta/\nu$ is expected to be
2 which is perfectly confirmed by our measurements: $\zeta/\nu=1.97$.
Looking at related systems, like doped germanium, indicates that the
observed behavior is a general one, but also that the actual exponents
depend on the doping range under consideration and the degree of
compensation.\cite{Ionov83,Watanabe00}

\section{Conclusions}

Our investigations of the electrodynamic properties of phosphorus doped
silicon in a broad frequency range from microwaves up to the far
infrared underline that electronic correlations between
the localized states play an important role at low temperatures.
The crossover in the optical conductivity from a linear to a quadratic
behavior as predicted by Efros and Shklovskii is observed
qualitatively; however, the simple model does not
lead to a quantitative agreement.
Our measurements show that the crossover energy $\hbar \omega_c$ is
not solely determined by the interaction energy $U(r_{\omega})$.
Furthermore, the Coulomb gap in the density of electronic states
explains the observation of a super-linear frequency dependence in the
Coulomb glass regime.
When approaching the MIT with increasing doping
concentration, the dielectric constant and the localization length show
critical behavior, in good agreement with theoretical
predictions.

However, there remain open questions to be addressed in the future.
Concerning the dielectric constant, there is the unclear origin of the
minimum in the temperature dependence. As far as the conductivity is concerned, our
results should be confirmed in comparable systems, like Si:B or Si:As. Of superior importance is a detailed investigation of the influence of compensation: while the theory of Shklovskii and Efros was done for compensated semiconductors, most experiments (ours as well) have been performed on nominally uncompensated samples.
Furthermore, studies in even larger parameter ranges are highly
desired to close the gap
not only in frequency but consequently also in concentration between
the GHz results of Lee and Stutzmann\cite{Lee2001} and the THz results
of the present and other recent studies.\cite{Helgren2002} Here the concentration dependence of the frequency crossover is of
particular interest.

Finally, it would be of interest to analyze the frequency-dependent conductivity at elevated temperature, i.e.\ when the zero-phonon regime is left. For $k_B T \approx \hbar\omega$ both photon-assisted and phonon-assisted processes are equally important. Although the data exist, suitable models are still lacking.

\begin{acknowledgments}
We would like to thank D. Schweitzer and W. Zulehner for providing the
single-crystalline Si of various doping concentrations. The samples
have been carefully prepared by G. Untereiner. We thank B. Gorshunov
for advice on measurement techniques. A.W. Anajoh performed some of the
dc experiments. We acknowledge the helpful discussions with G. Gr\"uner, E.
Helgren, M. Lee, E. Ritz and P. W\"olfle.
\end{acknowledgments}


\begin{thebibliography}{99}
\bibitem{remark4}One single crystal was grown by Soci\'et\'e G\'en\'erale M\'etal\-lurgique de Hoboken and provided by D.\ Schweitzer. The other set stems from Wacker Chemitronic and was provided by W.\ Zulehner; it was previously characterized in Karlsruhe using various techniques.\protect\cite{Lakner94,Gaymann93}
\bibitem{remark2}Indications of a mimimum in the temperature dependence of the dielectric constant can be found (without discussion) in the 3~kHz data of Tan and Castner\protect\cite{Tan81} on Si:P with $n=1.43\cdot10^{18}~{\rm cm}^{-3}$. No minimum is found for $n\leq 5.1\cdot 10^{17}~{\rm cm}^{-3}$.
\end{thebibliography}
\end{document}